% ...

\documentclass[floatfix,showpacs,amsmath,amssymb,aip,jmp,preprint]{revtex4-1}
\usepackage{overpic}
\usepackage{dsfont}
\usepackage{rotating}
\usepackage{xcolor}
\usepackage[urlcolor=blue,colorlinks=true,linkcolor=blue,pdfstartview={FitH},bookmarks=false]{hyperref}

\newcommand{\ket}[1]{\ensuremath{|#1\rangle}}

\newcommand{\braket}[2]{\ensuremath{\langle#1|#2\rangle}}

\newcommand{\eg}{\emph{e.g}}
\newcommand{\ie}{\emph{i.e}}

\newcommand{\mf}{\mathbf}

\newcommand{\smal}{\scriptscriptstyle}
\newcommand{\mc}{\mathcal}
\newcommand{\mb}{\mathbb}
\newcommand{\ot}{\otimes}

\newcommand{\idQ}{\mb{I}_{Q}}
\newcommand{\idE}{\mb{I}_{E}}

\newcommand{\p}{\scriptscriptstyle{+}}
\newcommand{\m}{\scriptscriptstyle{-}}
\newcommand{\spm}{\scriptscriptstyle{\pm}}

\newcommand{\txt}[1]{\text{#1}}

\begin{document}

\title{Notes on the Riccati operator equation in open quantum systems}
\date{18-04-2011}

\author{Bart{\l}omiej Gardas}
\email{bartek.gardas@gmail.com}

\affiliation{Institute of Physics, University of Silesia, PL-40-007 Katowice, Poland}

\author{Zbigniew Pucha{\l}a}
\email{z.puchala@iitis.pl}

\affiliation{Institute of Theoretical and Applied Informatics, Polish Academy of Sciences, Ba{\l}tycka 5, 44-100 Gliwice, Poland}

\begin{abstract}
A recent problem [B. Gardas, J. Math. Phys. {\bf 52}, 042104 (2011)] concerning an antilinear solution of the Riccati equation
is solved. We also exemplify that a simplification of the Riccati equation, even under reasonable assumptions, 
can lead to a not equivalent equation.
\end{abstract}
\maketitle



\section{Introduction} 
\label{sec:intro}

Recently, the Riccati operator equation (RE) has attached an attention to mathematical physicists (see \eg,
Refs.~\cite{gardas,*gardas2,*gardas3,*gardas4,gardasPuchala}).  As it has been argued, the solution of the
equation can be used to obtain reduced time evolution $\rho_t$ of a two-level open quantum system~\cite{breuer_book,davies}. This,
so-called the  reduced dynamics, plays a central role both in the quantum information theory and the theory of quantum
information processing. It is well-established that $\rho_t=\Phi_t(\rho_0)$, where $\Phi_t$ is the trace preserving (TP) and
completely positive (CP) map~\cite{choi,gorini}. Such TP-CP maps, known as channels, are commonly used to simulate real
quantum systems and to design quantum algorithms~\cite{Chuang,hayashi}.

To understand a connection between a given TP-CP map $\Phi_t$ and a solution of the RE, let us begin by reviewing some basics facts
concerning open quantum systems. For this purpose we consider the system-environment Hamiltonian in the following form
\begin{equation}
\label{total}
\mf{H} = H_{\text{Q}}\ot\idE+\idQ\ot H_{\text{E}} + \mf{H}_{\text{int}},
\end{equation}
where $H_{\text{Q}}$ and $H_{\text{E}}$ represent the Hamiltonian of the qubit and the environment, respectively. $\mf{H}_{\text{int}}$
specifies the interaction between the systems, whereas $\idQ$ and $\idE$ are the identity operators on corresponding Hilbert spaces 
$\mathbb{C}^2$ and $\mc{H}_{\text{E}}$ of the qubit and its environment, respectively. $\mf{H}$ acts on the Hilbert space 
$\mc{H}_{\text{tot}}=\mb{C}^{2}\ot\mc{H}_{\text{E}}$ and it admits the block operator matrix representation~\cite{spectral}:

\begin{equation}
 \label{bom}
  \mf{H} =
  \begin{bmatrix}
  H_{\p} & V \\
  V^{\dagger} & H_{\m}
  \end{bmatrix}
  \quad\text{on}\quad
  \mathcal{D}(\mf{H})=\mathcal{D}(H_{\p})\oplus\mathcal{D}(H_{\m}),
\end{equation}
with respect to the decomposition $\mc{H}_{\text{tot}}=\mc{H}_{\text{E}}\oplus\mc{H}_{\text{E}}$. The diagonal entries
$H_{\spm}:\mathcal{D}(H_{\spm})\rightarrow\mc{H}_{\text{E}}$ are densely defined and self-adjoint unbounded operators.
We have assumed that $\mc{D}(V)=\mc{D}(V^{\dagger})=\mc{H}_{\text{E}}$, which means that $V$, and thus $V^{\dagger}$ as
well, is bounded.

The time reduced evolution of an open system is given by
\begin{equation}
 \label{reduced}
 \Phi_t(\rho_0)  = \mbox{Tr}_{\text{E}}[\mf{U}_t\Psi(\rho_0)\mf{U}_t^{\dagger}],
\end{equation}
where $\mf{U}_t=\exp(-i\mf{H}t)$ is the time evolution operator of the total system. The map $\Psi$ assigns to each initial state
$\rho_0$ a single state $\Psi(\rho_0)$ of the total system. This assignment map must be chosen properly so that $\Phi_t$ can
be well-defined~\cite{reduced1,*reduced3,*reduced2}. In this paper we assume that no correlations between the systems are 
initially present~\cite{korelacje,*erratum}, and thus we take $\Psi(\rho_0)=\rho_0\otimes\omega$, for some initial state of
the environment $\omega$.

The linear map $\mbox{Tr}_{\text{E}}$ denotes the so-called partial trace:

\begin{equation}
 \label{Tr}
 \mbox{Tr}_E
 \begin{bmatrix}
  M_{11} & M_{12} \\
  M_{21}& M_{22}
  \end{bmatrix}
 =
 \begin{pmatrix}
  \mbox{Tr}M_{11} & \mbox{Tr}M_{12} \\
  \mbox{Tr}M_{21} & \mbox{Tr}M_{22}
  \end{pmatrix}
  \in M_2(\mathbb{C}),
  \quad\text{where}\quad M_{ij}\in\mathcal{T}(\mc{H}_{\text{E}}).
\end{equation} 
In the above description, $\mbox{Tr}$ refers to the usual trace on $\mc{H}_{\text{E}}$, 
$\mathcal{T}(\mc{H}_{\text{E}})$ denotes the Banach space of trace class operators with the trace norm: $\|A\|_{1}=\mbox{Tr}(\sqrt{AA^{\dagger}})$,
whereas $M_2(\mathbb{C})$ is the Banach space of $2\times 2$ complex matrices. Form~(\ref{Tr}) one can see that the partial trace in 
Eq.~(\ref{reduced}) transforms states of the total system (block operator matrices, square bracket) into states of the open system (complex matrices, 
round bracket). It is worth noting that the partial trace $\mbox{Tr}_{\text{E}}$, unlike the trace $\mbox{Tr}$, is not cyclic, namely 
$\mbox{Tr}_{\text{E}}(\mf{A}\mf{B})\not=\mbox{Tr}_{\text{E}}(\mf{B}\mf{A})$.

In general, the formula~(\ref{reduced}) is far less useful, than its simplicity might indicate. Indeed, to trace out the state 
$\mf{U}_t\Psi(\rho_0)\mf{U}_t^{\dagger}$ over the environment degrees of freedom one needs to find a decomposition of $\mf{U}_t$
with respect to the decomposition of Hilbert space $\mc{H}_{\text{tot}}=\mc{H}_{\text{E}}\oplus\mc{H}_{\text{E}}$.
This problem is fundamental in quantum mechanics, as it requires the diagonalization of $\mf{H}$. 

%
%
%

It is well-established (see \eg~\cite{Vadim} and Refs. therein) that $\mf{H}$ can be block-diagonalized 
provided there exist a bounded solution $X$ (with $\mbox{Ran}(X)_{|\mc{D}(H_{\p})}\subset\mc{D}(H_{\m})$) of
the Riccati equation:

\begin{equation}
 \label{ricc}
 XVX+XH_{\p}-H_{\m}X-V^{\dagger}=0\quad\text{on}\quad\mc{D}(H_{\p}).
\end{equation}
To be more specific, if $X$ solves~(\ref{ricc}) the following equality holds true: 
\begin{equation}
\label{de}
\mf{S}^{-1}\mf{H}\mf{S}
=
 \begin{bmatrix}
  Z_{\p} & 0 \\
  0 & Z_{\m}
  \end{bmatrix},
\quad\text{where}\quad
 \mf{S}=
 \begin{bmatrix}
  \idE & -X^{\dagger} \\
  X & \idE
  \end{bmatrix}
\end{equation}
and $Z_{\p}=H_{\p}+VX$, $Z_{\m}=H_{\m}-V^{\dagger}X^{\dagger}$ with $\mc{D}(Z_{\spm})=\mc{D}(H_{\spm})$. Moreover
$\mf{S}^{-1}$ exists as a bounded operator~\cite{gardasPuchala}. 

Using decomposition~(\ref{de}) one can write $\mf{U}_{t}$ in an explicit matrix form:

\begin{equation}
\label{class}
\mf{U}_t=
\exp\left(-i
\mf{S}
\begin{bmatrix}
  Z_{\p} & 0 \\
  0 & Z_{\m}
  \end{bmatrix}
  \mf{S}^{-1}t
\right)
=
\mf{S}
\begin{bmatrix}
  e^{-iZ_{\p}t} & 0 \\
  0 & e^{-iZ_{\m}t}
  \end{bmatrix}
  \mf{S}^{-1}.
\end{equation}

Among all difficulties concerning the RE, we will focus on two problems described bellow. The first one~\cite{gardas2}
arises when the solution of the RE is antilinear. In this case, the formula~(\ref{class}) cannot be applied, since $\mf{S}$
is neither an antilinear nor a linear operator. Therefore, we are interested in asking how can we determine the time evolution
of the total system.


The RE is an operator equation and even for a simple system it may be problematic to find the solution. 
However, there are cases in which the RE can be simplified considerably~\cite{gardas2,gardas3}. 
For instance, in the case when $V=\alpha\idE\equiv\alpha$, where $\alpha\in\mathbb{R}$; then the RE reads
\begin{equation}
 \label{sricc}
   \alpha X^2+XH_{\p}-H_{\m}X-\alpha = 0\quad\text{on}\quad\mc{D}(H_{\p}).
\end{equation}
%
In the instance when the operators $H_{\spm}$ are self-adjoint and commute (\eg, $[H_{\p},H_{\m}]=0$), it is reasonable to assume, that
the solution has the following form $X=f(H_{\p})$ or $X=g(H_{\m})$. Then, the functions $f$ and $g$ read $f=h_{|\sigma(H_{\p})}$
and $g=h_{|\sigma(H_{\m})}$, where $h$ satisfies the quadratic equation: $\alpha h(\lambda)^2+2h(\lambda)\lambda-\alpha=0$. At this
point the second issue can be addressed: does every solution of~(\ref{sricc}) is an analytical function of $H_{\p}$ or $H_{\m}$?

The primary goal of the presented work is to solve this two worthwhile issues. First, we show that there is a simple way of computing the
evolution operator $\mf{U}_t$ with the use of a formula similar to the one provided by the Eq.~(\ref{class}). Next, we exemplify that not
every solution of~(\ref{sricc}) is an analytical function of $H_{\p}$ or $H_{\m}$.

\section{Antilinear solution}
  \label{sol}
In this section we derive a simple method of computing $\mf{U}_t$ in the case when the solution of the RE is antilinear. To end this 
let us assume that $\tau$ is a bounded, antilinear (\ie, $\tau(z\ket{\psi})=z^*\tau\ket{\psi}$) operator which solves Eq.~(\ref{ricc})
and let $\mf{S}_{\tau}=\tfrac{1}{\sqrt{2}}\mf{S}$, where $\mf{S}$ denotes matrix from Eq.~(\ref{de}) with $X=\tau$. Although, 
$\mf{S}_{\tau}$ is neither a linear nor an antilinear in this case, it has an intersecting property:
\begin{equation}
\mf{S}_{\tau}(z\ket{\psi}) = \Re(z)\mf{S}_{\tau}\ket{\psi}+i\Im(z)\mf{S}_{{\smal{-}}\tau}\ket{\psi}
\quad\text{(note}\quad
\mf{S}_{{\smal{-}}\tau}\mf{S}_{\tau}=\idE=\mf{S}_{\tau}\mf{S}_{{\smal{-}}\tau}\text{)},
\end{equation}
which reduces to $\mf{S}_{\tau}(z\ket{\psi})=z\mf{S}_{\tau}\ket{\psi}$, for $z\in\mathbb{R}$. Therefore, $\mf{S}_{\tau}$ can be treated
as a linear operator when it acts on some subspace $\mc{H}_{\mathbb{R}}$ over a field of the \emph{real} numbers $\mb{R}$. 
In other words, the restricted operator $\mf{S}_{\tau|\mc{H}_{\mathbb{R}}}$ is linear.

As a result, if $\mf{H}=\mf{S}_{\tau}\mf{H}_{\txt{d}}\mf{S}_{{\smal{-}}\tau}$ then for an analytical function
$f:\sigma(\mf{H})\rightarrow\mathbb{R}$ we have $f(\mf{S}_{\tau}\mf{H}_{\txt{d}} \mf{S}_{{\smal{-}}\tau})=
\mf{S}_{\tau}f(\mf{H}_{\txt{d}})\mf{S}_{{\smal{-}}\tau}$, thus in particular we obtain
\begin{equation}
  \label{true}
    \exp(-i\mf{H}t) =   \mf{S}_{\tau}
    \cos(\mf{H}_{\txt{d}}t)\mf{S}_{{\smal{-}}\tau}
                        -i\mf{S}_{\tau}\sin(\mf{H}_\txt{d}t)\mf{S}_{{\smal{-}}\tau}.
\end{equation}
Although the right side of the Eq.~(\ref{true}) is \emph{not} equal to $\mf{S}_{\tau}\exp(-i\mf{H}_{\txt{d}}t)\mf{S}_{{\smal{-}}\tau}$,
yet the evolution operator can still easily be managed by separately transforming $\cos(\mf{H}t)$ and $\sin(\mf{H}t)$ functions, and then
combing the results by using the Euler formula.

\section{Non-analytical solution of the Riccati equation}

In this section we give an example that the solution of~(\ref{sricc}) is not necessarily an analytical function of $H_{\p}$ or $H_{\m}$.
To achieve this, we consider the RE~(\ref{sricc}) with $H_{\spm}=\pm H$, where $H:\mc{D}(H)\rightarrow\mc{H}_{\text{E}}$ is assumed to 
be densely defined and self-adjoin operator. With the use of these assumptions the RE on $\mc{D}(H)$ can be written as 
\begin{equation}
 \label{sricc2}
 \alpha X^2+XH+HX-\alpha = 0.
\end{equation}
%
If one assumes that the solution is in the form $X=f(H)$, then the RE~(\ref{sricc2}) simplifies on $\mc{D}(H)$ to the quadratic
operator equation:
\begin{equation}
 \label{sricc3}
 \alpha X^2+2HX-\alpha = 0.
\end{equation}
It is not difficult to see that the function $f$ is a solution of the quadratic equation:
\begin{equation}
 \alpha f(\lambda)^2+2\lambda f(\lambda)-\alpha = 0, \quad\text{where}\quad\lambda\in\sigma(H).
\end{equation}

Since there is only one self-adjoint operator in Eq.~(\ref{sricc2}), the assumption $X=f(H)$ seems to be reasonable. 
The question concerning analytical solutions of RE in this case reduces to the question: are the Eqs.~(\ref{sricc2})
and~(\ref{sricc3}) equivalent? We will exemplify that the answer in general is negative, and we will prove it
choosing $H$ such that the solution of~(\ref{sricc2}) does not solve~(\ref{sricc3}).

We construct $H$ as follows. Let $a^{\dagger}$ and $a$ be the bosonic creation and annihilation operators, respectively. They
are defined on a common domain $\mc{D}_1$ and obey the canonical commutation relation: $[a,a^{\dagger}]=\mb{I}$, on some dense
subset $\mc{D}_2$ of $\mc{D}_1$. In terms of this operators we define the number operator $N=a^{\dagger}a$, which is self-adjoint
on $\mc{D}_2$. An explicit form of the sets $\mc{D}_k$ is given by~\cite{sz0,*sz1,*sz2}

\begin{equation}
\mc{D}_k=\left\{\ket{\psi}: \sum_{n=0}^{\infty}n^k|\braket{\phi_n}{\psi}|^2<\infty\right\},\quad k=1,2;
\end{equation}
where the vectors $\ket{\phi_n}\in\mc{D}_2$ satisfy $N\ket{\phi_n}=n\ket{\phi_n}$. Moreover, on $\mc{D}_1$ the creation and annihilation
operators can be defined explicitly as 

\begin{equation}
 \label{aa}
   a\ket{\phi} = \sum_{n=1}^{\infty}\sqrt{n}\braket{\phi_n}{\phi}\ket{\phi_{n-1}}, \quad
   a^{\dagger}\ket{\phi} = \sum_{n=0}^{\infty}\sqrt{n+1}\braket{\phi_n}{\phi}\ket{\phi_{n+1}},
   \quad\ket{\phi}\in\mc{D}_1.
  \end{equation}
Finally, we take $H = g^*a+ga^{\dagger}$ with $\mc{D}(H)=\mc{D}_1$, where $g\in\mathbb{C}$. 

It is not difficult to see that the solution of the RE~(\ref{sricc2}) is given by the well-known bosonic parity operator~\cite{Bender}:

\begin{equation}
 \label{parity}
 P\ket{\psi} = \sum_{n=0}^{\infty}e^{i\pi n}\braket{\phi_n}{\psi}\ket{\phi_n},\quad\ket{\psi}\in\mc{H}_{\text{E}}.
\end{equation}
Indeed, from~(\ref{parity}) follows that $P$ is both Hermitian and unitary, in particular $P^2=\idE$. Therefore, in order to prove
that $P$ solves~(\ref{sricc2}), it is sufficient to show that $PHP=-H$. The latter is obvious since $PaP=-a$ and $Pa^{\dagger}P = 
-a^{\dagger}$.

Note, formally $P$ can be written as $P=\exp(i\pi N)$, however, unlike $N$ it is everywhere defined, thus it is bounded
as well ($\|P\|=1$). Of course, $P$ is not an analytical function of $H$ and it does not solve Eq.~(\ref{sricc3}) because
$PH\not=0$.

\section{Summary}

We have proposed a simple resolution of the recent problem concerning antilinear solution of the Riccati
equation. Our result allows to take advantage of the result~(\ref{class}), in the case of an antilinear decomposition
of the total Hamiltonian. The solution given in this paper is not as complicate as it is crafty. It relays on the very
same exponential formula that is uses for a linear decomposition of the total Hamiltonian. However, the way it is used
is different as we have explained in Sec.~\ref{sol}.

We have also argued that a simplification of the RE, even under reasonable assumptions, can lead to the not equivalent
equation.

\section{Acknowledgments}
B. Gardas would like to acknowledge a scholarship from the TWING project co-finance
by the European Social Fund. Work by Z. Pucha\l{}a was financed from the resources of
National Science Centre as a research project - grant number N~N514~513340.

%
%

\end{document}